\documentclass[conference,9pt]{IEEEtran}
\IEEEoverridecommandlockouts
\usepackage{cite}
\usepackage{amsmath,amssymb,amsfonts}
\usepackage[ruled, vlined, linesnumbered]{algorithm2e}
\usepackage{graphicx}
\usepackage{textcomp}
\usepackage{xcolor}
\usepackage{textcase}
\usepackage{hyperref}
\usepackage[tablename=TABLE]{caption}
\DeclareCaptionTextFormat{up}{\MakeTextLowercase{#1}}
\newenvironment{myitemize}{\begin{list}{$\bullet$}
{\setlength{\topsep}{1mm}
\setlength{\itemsep}{0.25mm}
\setlength{\parsep}{0.25mm}
\setlength{\itemindent}{0mm}
\setlength{\partopsep}{0mm}
\setlength{\labelwidth}{15mm}
\setlength{\leftmargin}{4mm}}}{\end{list}}

\SetKwInput{KwInput}{Input}                %
\SetKwInput{KwOutput}{Output}

\newtheorem{definition}{Definition}

\newtheorem{remark}{Remark}
\newcommand{\FS}[2]{\displaystyle\frac{#1}{#2}}
\DeclareMathOperator*{\argmax}{arg\,max}
\DeclareMathOperator*{\argmin}{arg\,min}

\definecolor{mypink}{cmyk}{0, 0.7808, 0.4429, 0.1412}

\newcommand{\toolname}{\textsc{Cocktail}}

\def\BibTeX{{\rm B\kern-.05em{\sc i\kern-.025em b}\kern-.08em
    T\kern-.1667em\lower.7ex\hbox{E}\kern-.125emX}}
\begin{document}

\title{Cocktail: Learn a Better Neural Network Controller from Multiple Experts via Adaptive Mixing and Robust Distillation }

\author{
\IEEEauthorblockN{Yixuan Wang,
Chao Huang,
Zhilu Wang,
Shichao Xu, 
Zhaoran Wang,
Qi Zhu}
\IEEEauthorblockA{Northwestern University, Evanston, IL\\
\{yixuanwang2024@u., chao.huang@, zhilu.wang@u., shichaoxu2023@u., qzhu@\}northwestern.edu, zhaoranwang@gmail.com }
}

\maketitle

\begin{abstract}

Neural networks are being increasingly applied to control and decision making for learning-enabled cyber-physical systems (LE-CPSs). They have shown promising performance without requiring the development of complex physical models; however, their adoption is significantly hindered by the concerns on their safety, robustness, and efficiency. In this work, we propose \toolname, a novel design framework that automatically learns a neural network based controller from multiple existing control methods (experts) that could be either model-based or neural network based. In particular, \toolname \ first performs reinforcement learning to learn an optimal system-level adaptive mixing strategy that incorporates the underlying experts with dynamically-assigned weights, and then conducts a teacher-student distillation with probabilistic adversarial training and regularization to synthesize a student neural network controller with improved control robustness (measured by a safe control rate metric with respect to adversarial attacks or measurement noises), control energy efficiency, and verifiability (measured by the computation time for verification).    
Experiments on three non-linear systems demonstrate significant advantages of our approach on these properties over various baseline methods.

\end{abstract}


\section{Introduction}

Machine learning techniques, particularly those based on neural networks, have seen rapidly growing applications in autonomous cyber-physical systems such as self-driving vehicles, smart buildings, and robotic systems. These learning-enabled cyber-physical systems (LE-CPSs) adopt machine learning techniques not only for perception of the environment~\cite{redmon2016you}, but increasingly also for  control~\cite{yang2020rigid} and decision making, in large part due to their advantages in learning effective strategies without the need of developing complex, costly, and error-prone physical models~\cite{xu2020one}.   
However, applying neural networks for building autonomous CPSs still faces significant hurdles, particularly with concerns of their impact on system safety, robustness, and efficiency. To enable their wider adoption, it is important to develop automated design methods and tools for analyzing these properties and optimizing the control design accordingly.

In this paper, we present \toolname, 
a novel framework for learning an improved neural network controller from multiple existing control methods (or so-called ``experts''). This is based on the observation that for many control applications, there are often multiple candidate control methods (experts) available~\cite{wang2020energy}. They could be based on well-established model-based approaches, such as model-predictive control (MPC)~\cite{qin2003survey} or linear quadratic regulator (LQR)~\cite{bemporad2002explicit}. They could also be neural network based control methods that are trained through different algorithms, e.g,. via various reinforcement learning (RL) approaches, and have different architectures and hyper-parameters. In practice, it is also common for LE-CPSs to have multiple available controllers that are designed by different teams and/or for different objectives. Note, our framework does not require the experts to be optimal.  

The multiple available controllers/experts, which may include both model-based and neural network-based ones, often perform differently and have different strengths with respect to the changing system state. Thus, the first step of our framework \toolname \ is to learn a system-level \textbf{adaptive mixing} strategy that linearly combines the multiple available experts with dynamically-assigned weights for generating control input to the system. The weights are adapted based on the system state at each sampling period, to optimize system control robustness and control energy efficiency. Note that the robustness objective is defined as a safe control rate metric (i.e., how likely the system can remain safe from any initial state) under optimized adversarial attacks or random measurement noises to the system state.
We formulate this adaptive mixing problem as a Markov Decision Process (MDP) with a reward function modeling robustness and efficiency, and ensure the optimality of our RL-based solution with global optimum convergence analysis~\cite{liu2019neural}. 

While the adaptive mixing strategy can leverage the strengths from multiple experts and effectively improve the control robustness and energy efficiency, the mixed controller design could take significant resources (e.g, in storage) to implement and very importantly, be difficult to formally verify its properties such as safety and robustness. Thus, the second step of \toolname \ conducts a teacher-student \textbf{robust distillation} to synthesize a single student neural network from the mixed controller design, using a novel probabilistic adversarial training and regularization technique with dual-objective regression focusing on both robustness and verifiability (measured by the computation time for verification). As we observed in experiments, this provides significant further improvement on all the properties we consider, including robustness, verifiability, and energy efficiency.


 \smallskip
 \noindent
 \textbf{Related work:} Our work is related to a rich literature on adaptive controller design. For instance, simplex architecture~\cite{seto1998simplex} proposes a switching logic between a baseline controller and an advanced controller to improve the control performance. Control adaptation based on switching among multiple controllers/experts has also been addressed in~\cite{gong2004heuristic} with a rule-based approach, in~\cite{huang2020opportunistic, wang2020energy} with an RL approach for energy efficiency, and in~\cite{ramakrishna2020dynamic} with finite-size weighted adaptation based on Q-learning. Different from these discrete adaptation approaches, we consider a continuous version of adaptive mixing, whose feasible adaptation space is a super-space of the ones in these previous approaches. We find that by expanding the adaptation space, our approach can significantly improve the safe control rate over the literature (as reported later in our experiments). Moreover, we dynamically optimize the weights with global convergence assurance, which is not guaranteed in the literature.

 

Our work also relates to the knowledge distillation paradigm~\cite{hinton2015distilling}, where a complex neural network is distilled into a compact neural network with similar or even better performance. Distillation from multiple experts, i.e., an ensemble of teachers, has been considered in works such as~\cite{fukuda2017efficient,chebotar2016distilling}. In these approaches, the weight for each teacher in the ensemble is pre-determined and the sum of the weights is constrained to 1. In contrast, our approach dynamically adjusts the weights with RL, and does not put constraint on the weight sum to facilitate the implementation of the RL process. Moreover, our distillation is based on a novel dual-objective process with consideration of both robustness and verifiability. 


\noindent
In summary, our work makes the following contributions:

\begin{myitemize}
\item We propose the \toolname \ framework to leverage multiple existing control methods (experts) and learn a better single neural network controller from them, with consideration of control robustness, control energy efficiency, and verifiability.
\item The \toolname \ framework includes two novel components. The adaptive mixing step uses RL to learn a system-level strategy for dynamically assigning weights in incorporating experts, with global optimum convergence assurance. The robust distillation step conducts probabilistic adversarial training and regularization to synthesize a single neural network controller that further improves the mixed controller design. 
\item Experiments on three non-linear systems demonstrate that our approach can significantly improve robustness, energy efficiency, and verifiability over various baseline methods, including any single expert and a state-of-the-art switching adaptation method from the literature.
\end{myitemize}
 
In the rest of the paper, Section~\ref{problem_formulation} presents the problem formulation. Section~\ref{our_approach} presents our \toolname \ framework. Section \ref{experiment} shows the experimental results, and Section~\ref{conclusion} concludes the paper.

\section{Problem Formulation}
\label{problem_formulation}

We consider a discrete-time feedback system with its dynamics as
\begin{equation}
    s(t+1) = f(s(t), u(t), \omega(t), \delta(t)), \ \forall t \geq 0
    \label{dynamics}
\end{equation}
where $f:\mathbb{R}^{|s|} \times \mathbb{R}^{|u|} \times \mathbb{R}^{|\omega|} \times \mathbb{R}^{|\delta|} \rightarrow \mathbb{R}^{|s|}$ is a locally Lipschitz-continuous function~\cite{ruan2018reachability}.
 $s(t)  \in \mathbb{R}^{|s|}$ is the system state vector. $X$ is defined as the \textit{safe region}, and any state out of $X$ is considered unsafe. $X_{0} \subseteq X$ is the set of all possible initial system states. $u(t) \in U \in \mathbb{R}^{|u|}$ is the feedback control input to the system plant at each timestep $t$, where $U$ is the bound for vector $u(t)$. $\omega(t) \in \Omega \in \mathbb{R}^{|\omega|}$ is a bounded external disturbance. $\delta(t) \in \Delta$ is a perturbation to the system state that could be caused by targeted/optimized adversarial attacks or random measurement noises. Note that $X$, $X_{0}$, $U$, $\Omega$, and $\Delta$ are constrained by pre-defined functions, such as boxes. 

The above system can be controlled with a feedback controller $\kappa$ that is either model-based or model-free (e.g., those based on neural networks). At each timestep $t$, the controller $\kappa$ reads the system state $s(t)$, and computes a control input as $u(t) = \kappa(s(t))$. The system then evolves to $s(t+1)$ according to its dynamics in Eq~\eqref{dynamics}. Such process repeats and a trajectory $\varphi$ based on the system initial state $s(0) \in X_{0}$ and the controller $\kappa$ can be defined as
\begin{equation}
    \varphi_{s(0), \kappa}(t+1) = f(\varphi_{s(0), \kappa}(t), \kappa(\varphi_{s(0), \kappa}(t)), \omega(t), \delta(t))
\end{equation}

A trajectory is safe if every state it visits is within the safe region $X$. For a controller $\kappa$, we can define a \textit{safe initial state set} $X^{'}$, which includes any initial state whose trajectory under $\kappa$ is safe, i.e., 
\begin{displaymath}
X^{'}_{\kappa} = \{s \ | \ s\in X_{0}, \varphi_{s, \kappa}(t) \in X, \ \forall t \geq  0\}
\end{displaymath}
We can then define a \textit{safe control rate} metric for each controller $\kappa$ to measure how large its safe initial state set $X^{'}_{\kappa}$ is, with respect to the set of all possible initial states $X_0$ (i.e., the ratio between the sizes of the two sets). 

Based the above system model, we define three properties for a controller $\kappa$ as follows.

\textit{Property 1:} \textbf{Control robustness} for a controller $\kappa$ is defined as its safe control rate $S_r$ under optimized adversarial attacks or random measurement noises on the system state (captured by the state perturbation $\delta(t)$). Note that system \textit{safety} may be considered as a special case of robustness with $0$ state perturbation.  

\textit{Property 2:} \textbf{Control energy efficiency}~\cite{wang2020energy} for a controller $\kappa$ is defined as the average control energy cost $e$ (over $T$ control steps) of the various trajectories generated from the initial states in its safe initial state set $X'_{\kappa}$, i.e., 
\begin{equation}\label{energy_cost}
e = \mathbb{E}\left[\sum_{t=0}^{T-1} ||\kappa(\varphi_{s,\kappa}(t))||_{1}\right], \forall s \in X^{'}_{\kappa}
\end{equation}
where $\left\|\cdot \right\|_{1}$ is the 1-norm operator. This metric is evaluated via sampling of the initial state set in our experiments.

\textit{Property 3: }\textbf{Verifiability} is measured by the computation time of the verification processes for various properties on a given platform.





The problem we try to solve is then defined as: given a system as described in Eq~\eqref{dynamics} and multiple control experts $\kappa_{i} (i=1,\cdots, n)$(not necessary to be optimal), we will design a new neural network controller $\kappa^{*}$ that optimizes control robustness, control energy efficiency, and verifiability.


              




\section{Our \toolname \ Framework}
\label{our_approach}

\begin{figure}[ht]
    \centering
    \includegraphics[width=1.0\linewidth]{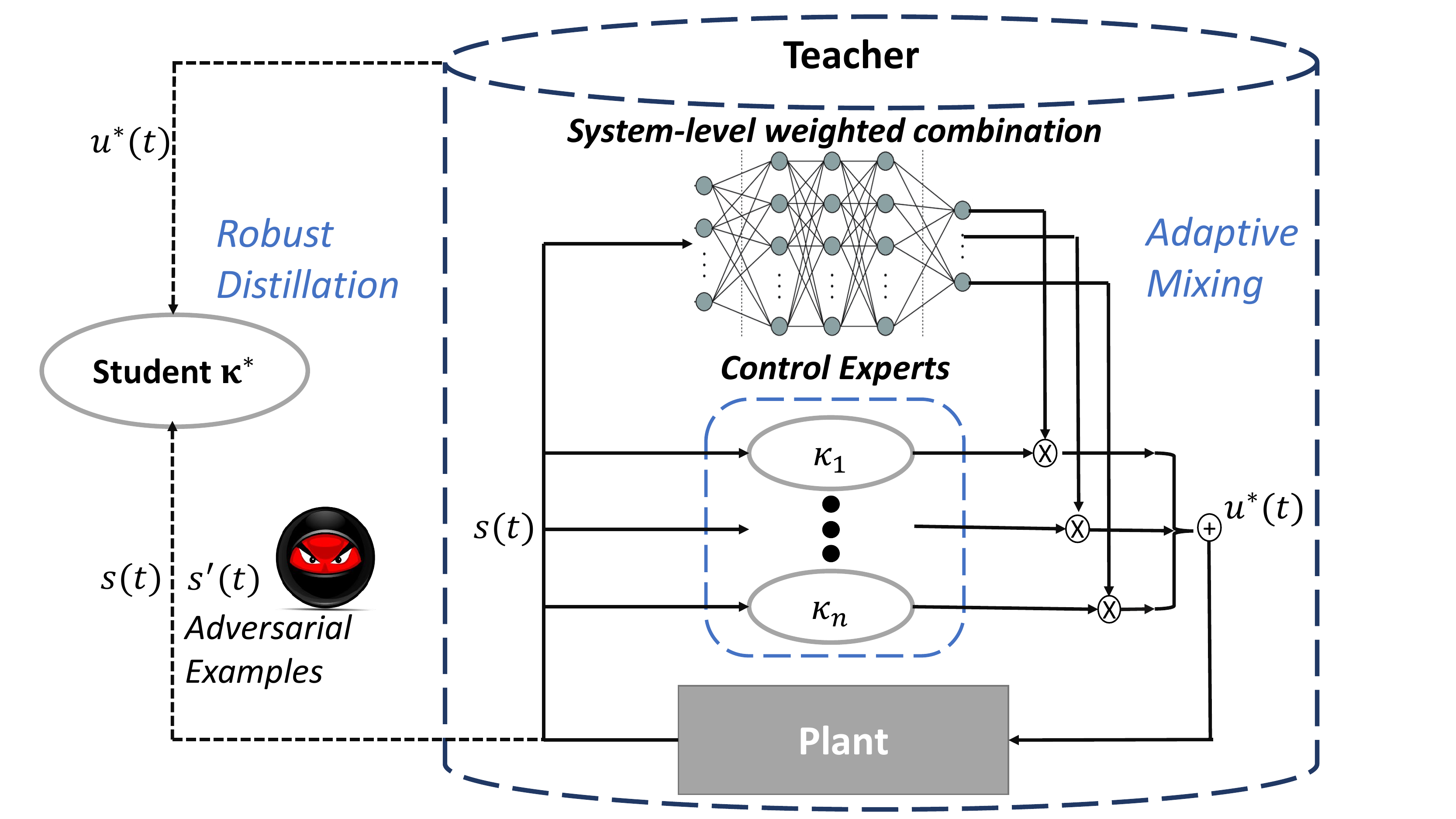}
    \caption{Overview of the proposed \toolname \ framework. 
    }
    \label{overview_fig}
    \vspace{-6pt}
\end{figure}

This section presents our proposed \toolname \ framework for solving the above problem defined in Section~\ref{problem_formulation}. As shown in Fig.~\ref{overview_fig}, the \toolname \ framework includes two novel components. First, a system-level adaptive mixing strategy linearly combines the multiple control experts for generating the control input to the system plant. The weights for the linear combination are dynamically adapted based on the system state, and learned via RL according to an MDP formulation that optimizes control robustness (i.e., safe control rate) and energy efficiency with global optimum assurance. 
 Then, through teacher-student knowledge distillation, a student neural network $\kappa^{*}$ is learned from the mixed controller design (which includes the underlying experts and the system-level neural network learned via RL for generating the weights). The distillation process is based on a probabilistic adversarial training and regulation technique that further improves control robustness and verifiability via minimizing the Lipschitz constant of the student network. 
Once we obtain the distilled student controller, various formal verification techniques can be effectively applied to analyze its properties such as safety. 
More details of \toolname \ is shown in Algorithm~\ref{alg} and introduced in the remaining of the section. 

\subsection{RL-based Adaptive Mixing of Multiple Experts}

We propose to learn a system-level adaptive mixing strategy that significantly expands the action/adaptation space of the switching control methods in the literature (e.g., those in~\cite{seto1998simplex, wang2020energy}). In principle, we could build any mapping function $g:\mathbb{R}^{n\times |u|} \rightarrow \mathbb{R}^{|u|}$ that maps the various control input values computed by the experts to a control input for the system plant. In this work, we focus on linear mapping functions and dynamically adjust the weights for each expert based on the system state. 
 To achieve this, we formulate the learning process for such adaptive mixing strategy as an MDP and solved with RL, with control robustness and energy efficiency as the reward. 
 
  Our MDP is captured with a tuple ($\mathcal{S, A, P, R, \gamma}$). $\mathcal{S}$ is the system state space, and $\mathcal{A}$ is the action space. $\mathcal{P}:\mathcal{S} \times \mathcal{A} \rightarrow \mathcal{S}$ describes the state transition function or system dynamics, which is invisible to the controllers. $\mathcal{\gamma} \in (0, 1]$ is a constant discount factor. Parameterized by $\theta$,  policy $\pi_{\theta} \in \Pi:\mathcal{S} \rightarrow \mathcal{A}$ denotes the strategy. More specifically, they are formulated as follows.
 
 \begin{algorithm}
\caption{Proposed \toolname \ Framework} 
\label{alg} 
\SetAlgoLined
\DontPrintSemicolon
\KwInput {Multiple control experts $\kappa_{i}, i = 1, \cdots, n$}
\KwOutput{Student controller $\kappa^{*}(;q)$}

Initialize replay memory $D$, adaptive policy network $\pi_\theta$, state perturbation bound $\Delta$, epochs $N$, steps  $T$, Distillation epoch $N_E$, weights $\beta, \lambda$ and probability $p$.

\For{$epoch = 0, \ldots, N$} {
 Randomly initialize state $s(0) \in X_0, \theta_{old} \leftarrow \theta$.

\For {$t = 0, \ldots, T$} {

$a(t) = \pi_{\theta_{old}}(s(t))$. \\
\hspace*{-0.06in}$ u(t) = clip(\sum_{i=1}^{n} a(t)_{i}\times\kappa_{i}(s(t)), U_{inf}, U_{sup})$; \\ \hspace*{-0.06in}$s(t+1) = f(s(t), u(t), \omega(t), \delta(t))$; Receives $r(t)$; \\ 
\hspace*{-0.06in}$D.append([s(t), a(t), s(t+1), r(t)])$. \\

\tcc{RL(PPO) for adaptive mixing}
Sample mini-batch from $D$; Compute advantage function $\hat{A}$ \\ 

$\theta = \argmax\limits_{\theta} \mathbb{\hat{E}} \left[\FS{\pi_{\theta}(a | s)}{\pi_{\theta_{old}}(a | s)} \hat{A} - \beta \mathcal{KL}[\pi_{\theta_{old}}(\cdot| s), \pi_{\theta}(\cdot| s)] \right]$. \\
 
\tcc{Robust distillation}
\If{$epoch \geq N_E$}{
$z \xleftarrow[random]{uniform} [0, 1]$. 

$\delta = \Delta*sign(\nabla_{s}(l(\kappa^{*}(s;q), u)))$ if $z\leq p$ else 0. 

$q = \argmin \limits_{q}  l (\kappa^{*}(s+
    \delta;q), u) + \lambda  ||q||^{2}_{2}$
}
}
}
\end{algorithm}

 \smallskip
 \noindent 
 \textbf{State:} $\mathcal{S}$ is the system state space. In this paper, we assume that each $s \in \mathcal{S}$ can be observed but may be maliciously attacked or affected by random measurement noises. The attacks or noises are captured by a bounded perturbation $\delta$ to the system state as introduced in Section~\ref{problem_formulation}, and their effects reflect the control robustness (this will be detailed more later).
 
 
 \smallskip
 \noindent
 \textbf{Action:} We consider a linear mapping function in this paper to generate the action space $\mathcal{A}$ for our adaptive mixing strategy. Specifically, at each timestep $t$, the action $a(t)=(a_1, \cdots, a_n)$ 
 represents the weight assignment to the experts in the linear mapping function, where $a_i$ is a bounded weight assigned to the $i$-th expert ($a_i \in [-A_{B_i}, A_{B_i}]$, $A_{B_i} \geq 1$). 
Then, the control input to the system is the weighted sum of the control inputs computed by the experts, with a clipping function ensuring its feasibility:
 \begin{equation}
 u(t) = clip(\sum_{i=1}^{n} a(t)_{i}\times \kappa_{i}(s(t)), U_{inf}, U_{sup})
 \label{control_input}
 \end{equation}
 where $\kappa_{i}(s(t))$ is the control input value computed by the $i$-th expert. $U_{inf}$ and $U_{sup}$ are the infimum and supremum of the control input vector bound $U$, respectively. Note that as a polyhedron, the action space in our approach is a super-space of the one in~\cite{ramakrishna2020dynamic} (convex hull) and in~\cite{wang2020energy, huang2020opportunistic} (switching). 
 
 \smallskip 
 \noindent
 \textbf{Reward function:} The reward function encodes our desired goal for optimizing control robustness (i.e., safe control rate) and control energy efficiency, by steering the system away from the unsafe region and using as little energy as possible. Specifically, it is defined as
 \begin{displaymath}
 r(s, a) = \begin{cases}
  R_{pun}, \ \ \text{$if \ s \notin X$}\\
 h(||u||), \ otherwise
\end{cases}
 \end{displaymath}
 where $R_{pun}$ is a large negative punishment on safety violations (i.e., $s \notin X$). $h$ is a monotonically decreasing function that computes energy consumption based on the control input $||u||$ in Eq~\eqref{control_input}.
 
 With above design of the reward function, we formulate an optimization problem concerning robustness and efficiency as 
 \begin{displaymath}
 \begin{aligned}
 \max J_{\pi_{\theta}} &= \sum_{t=0}^{T-1} \mathbb{E} \left[\gamma^{t} \cdot r(s(t), a(t))\right] \\
 s.t. \  &  s(t+1)=f(s(t), u(t), \omega(t), \delta(t)), s(0) \in X_0 \\
 & a(t) = \pi_{\theta}(s(t)) \\
 &  -A_{B_i} \leq  a(t)_{i}  \leq A_{B_i}, \forall \ i=1, \cdots, n 
 \end{aligned}
 \end{displaymath}
 where $T$ is an episodic control length. 
 
 For each iteration in the learning of the adaptive mixing strategy in Algorithm~\ref{alg}, we solve the above optimization problem with the gradient ascent towards the optimal weights for the experts, i.e.,
 \begin{displaymath}
 \theta = \argmax\limits_{\theta} \mathbb{\hat{E}} \left[\FS{\pi_{\theta}(a | s)}{\pi_{\theta_{old}}(a | s)} \hat{A} - \beta \mathcal{KL}[\pi_{\theta_{old}}(\cdot| s), \pi_{\theta}(\cdot| s)] \right]
 \end{displaymath}
where $\hat{A}$ is the advantage function in RL, $\mathcal{KL}$ is the KL divergence, $\theta_{old}$ represents the parameters for the adaptive mixing policy network from the last iteration, and $\mathbb{\hat{E}}$ is an estimator (sample mean) for the expectation. Our approach can converge to the optimal weight assignment for the optimization problem, as explained below.

\newtheorem{prop}{Proposition}
\begin{prop}
Given multiple experts $\kappa_{i} (i=1, \cdots, n)$, our RL-based approach can learn an optimal policy $\pi^{*}$ for the adaptive weight assignment of experts, and outperform (or perform equally to) any single expert controller or any switching adaptation policy $\pi_{s}$. 
\end{prop}


\textit{Proof:} First, according to~\cite{liu2019neural}, the actor-critic methods for proximal policy optimization (PPO)~\cite{schulman2017proximal} with neural networks approximation converge to the global optimum at a sub-linear rate. This applies to our approach.
Moreover, the action space of any switching adaptation policy that switches among controllers (e.g., the one in~\cite{wang2020energy}) or of any policy with finite-size weighted adaptation (e.g., the one in~\cite{ramakrishna2020dynamic}) is a sub-space of our action space. As global optimum is better than or equal to any local optimum, the optimal policy $\pi^{*}$ obtained in our approach should outperform or perform equally to the ones from any single expert or switching policy.



\begin{remark}
The optimality assurance only applies to PPO in principle~\cite{liu2019neural}. In practice, however, we find that other RL methods such as the deep deterministic policy gradient (DDPG)~\cite{lillicrap2016continuous} can also achieve significant improvement. 

\end{remark}


\subsection{Robust Distillation to a Single Neural Network Controller} 

The adaptive mixing strategy can effectively leverage the strengths from multiple experts to improve control robustness and energy efficiency. However, the learned mixed controller design, with the multiple underlying experts and a neural network for the adaptive mixing policy, may consume significant resources in implementation (e.g., large storage requirement). Moreover, it is hard to formally verify the properties for such mixed controller due to its complexity. This motivates us to further synthesize a single and simpler neural network controller via knowledge distillation. 

An important observation that drives our distillation is that for a neural network, both its verification complexity and its robustness are often affected by its Lipschitz constant $L$\footnote{The Lipschitz constant of a layer that is parameterized by weights $\mathcal{W}$ in a fully-connected feed-forward neural network with activation functions $Relu$, $Tanh$ and $Sigmoid$ can be computed as $||\mathcal{W}||$, $||\mathcal{W}||$ and $\frac{1}{4} ||\mathcal{W}||$, respectively. The overall Lipschitz constant of the neural network is the product of each layer's.}. Typically, the smaller the Lipschitz constant is, the more robust and more verifiable (e.g., taking less time to verify certain properties) the neural network is~\cite{fan2019towards, pauli2020training}.

Thus, the goals for our distillation of the student network are two folds: 1) to achieve similar control performance (in this case the control energy efficiency) as the mixed controller design (i.e, the teacher), by minimizing a loss function that measures the regression error between the student and the teacher; and 2) to further improve system verifiability and control robustness via minimizing the Lipschitz constant of the student network. 

To achieve our dual objectives, we propose a hybrid probabilistic learning process by randomly selecting direct distillation or adversarial training with the fast-gradient sign method (FGSM)~\cite{goodfellow2014explaining} and L-2 regularization to reduce $L$, as shown in Algorithm~\ref{alg}.
Specifically, the  part of the adversarial training with regulation solves a min-max problem each time as:  
\begin{displaymath}
\min \limits_{q}(\max \limits_{||\delta|| \leq \Delta} l(\kappa^{*}(s+\delta;q), u) + \lambda ||q||^{2}_{2})
\end{displaymath}
where $\kappa^{*}$ is the distilled student network with parameters $q$. $\delta$ bounded by $\Delta$ is the perturbation on the system state, which may be caused by adversarial attacks or measurement noises. $l$ is a loss function that measures the regression error between the student network and the teacher in mean squared error (MSE), and $\lambda$ is the weight for the regularization.
Intuitively, minimizing this training loss will regulate the local Lipschitz constant, as the output of neighbour region of $s$ is expected to map closed to $u$.  
The inner max problem is solved by adversarial example generation with gradient ascent method and sign function as
\begin{displaymath}
\delta = \Delta*sign(\nabla_{s}(l(\kappa^{*}(s;q), u)))
\end{displaymath}

Through this min-max optimization, the Lipschitz constant of the distilled student network can be significantly reduced, improving both system verifiability and control robustness; while similar energy efficiency can be achieved (in experiments, it is actually also improved).




\subsection{Verification of the Neural Network based Controllers}
\label{verification}

Once we obtain the distilled student neural network $\kappa^{*}$, we may formally evaluate some of its properties such as safety and robustness, using techniques such as control invariant set computation and reachability analysis for safety verification 
(recall that our robustness property is defined based on the safe control rate under attacks or measurement noises to the system state). 
Intuitively, a control invariant set is a subset of the safe region that every possible trajectory starting from it will never leave it. To compute the invariant set, reachable analysis is used to compute the set (or an over-approximation of it) of all possible states the system may visit within a finite-horizon timestep. They are
more formally defined as follows.
\begin{definition}
A control invariant set $X_{I}$ is a subset of the safe region $X$ that is defined as
\begin{displaymath}
X_{I} = \{s \ | \ \varphi_{s, \kappa}(t) \in X_{I} \in X, \ \forall t \geq 0, \ \forall \omega(t) \in \Omega \}
\end{displaymath}
\end{definition}

Note that any initial state within the invariant set is guaranteed to have infinite-time horizon safety as its possible trajectories are bounded within the invariant set.

\begin{definition}
The reachable set for an initial state $s_{0} \in X_{0}$ is the set of states that the system may reach within $T$ timesteps, i.e.,
\begin{displaymath}
X_{R} = \{ \varphi_{s_{0}, \kappa}(t) \ | \ \forall \ s(0) \in X_{0},  \ \forall \  0 \leq t \leq T-1 \}
\end{displaymath}
\end{definition}

Directly performing reachability analysis and safety verification on neural networks is intractable in most cases. Thus, we leverage the methods from~\cite{wang2020energy, huang2019reachnn} by first over-approximating the neural network controller with a Bernstein polynomial under bounded errors (with partitioning technique~\cite{huang2019reachnn} for reducing the approximation error), and then transforming the entire system (including the plant) into a hybrid system. The system safety and the robustness property (safe control rate under attacks or noises) can then be evaluated on the hybrid system with existing tools from~\cite{xue2018robust, chen2013flow}.
Specifically, in mathematical form, we first approximate the student network $\kappa$ with a Bernstein polynomial as follows:
\begin{displaymath}
 \kappa^{*}(x) \in B_{d}(x) +  [-{\epsilon}, {\epsilon}],\ \forall x \in X
\end{displaymath}
where $d$ is the degree of the Bernstein polynomial and $\epsilon$ is the absolute approximation error bound. 
If the approximation error is too large, we can further partition the system state  as:
\begin{displaymath}
 \kappa^{*}(x) \in B_{d}^{p}(x) +  [-\hat{\epsilon}^{p}, \hat{\epsilon}^{p}],\ \forall x \in X^{p}, \forall p = 1, \cdots, P.
\end{displaymath}
where $P$ is the number of partitions and $\epsilon = \max(\hat{\epsilon}^{p})$ is the approximation error. Such error will eventually be counted as an additional external disturbance into the original system as $\hat{\Omega} = \Omega \bigoplus \epsilon$, where $\bigoplus$ is the Minkowski summation operator.

\begin{remark}
  Benefited from the robust distillation, the neural network controller $\kappa^{*}$ generated by \toolname \ with reduced Lipschitz constant is much more computationally efficient for verification purpose, compared with not only the mixed controller design (which is hard to verify with current tools) but also the student network generated from direct distillation (i.e., without adversarial training and regulation for reducing Lipschitz constant). 
  This is due to the fact that larger Lipschitz constant leads to more sampling, more partitions, and higher order of Bernstein polynomials for approximating the neural network. Moreover, the transformed hybrid system also has more optimization variables and requires more resources to verify. Note, although not testes in this paper, large Lipschitz constant of neural network controller is also expected to cause a significant impact on Verisig~\cite{ivanov2019verisig, fan2019towards}. 
\end{remark}

However, while system safety under no attack or measure noise can be effectively verified for our test examples using the generated student neural network (and demonstrated in our experiments), accurately computing the control robustness under attacks and noises is still quite challenging with the current formal analysis techniques, as the over-approximation error cannot be effectively reduced within reasonable computation time in this case~\cite{huang2019reachnn}. Thus, in our experiments, the safe control rate metric (i.e., robustness) for a controller is \emph{estimated} by picking random samples from the initial state set $X_0$ and evaluating the system safety under the controller via simulations. This is also because the safety for some baselines methods cannot be formally analyzed in any case with the current tools.

\section{Experimental Results}\label{experiment}

\noindent
\smallskip
\textbf{Test Systems:}  We conduct experiments on three non-linear systems: a Van der Pol's oscillator, a three-dimensional system from~\cite{sassi2017linear} (example 15), and a cartpole system. 
Each system has two available control experts $\kappa_1$ and $\kappa_2$, obtained by DDPG with different hyper-parameters, or in the case of the 3D system, DDPG and a model-based controller from~\cite{sassi2017linear}. More details are as follows.


1) The Van der Pol's oscillator is described as 
\begin{equation} \label{OS_dynamics}
\begin{cases}
s_{1}(t{+}1) =s_{1}(t) + \tau s_{2}(t)\\
s_{2}(t{+}1) =s_{2}(t) + \tau[(1{-}s_{1}^{2}(t))s_{2}(t) {-} s_{1}(t) {+} u(t)] + \omega(t)
\end{cases}
\end{equation}
where $s(t) = (s_{1}(t), s_{2}(t))^{'}$ is the system state.  $X = X_{0} = [-2, 2]^{2}$ (for further control invariant analysis). $u(t)$ is the control input variable, and is bounded by $[-20, 20]$. External disturbance $\omega$ is a random variable uniformly sampled from $[-0.05, 0.05]$. $\tau=0.05$ is the sampling period. We assume that each control epoch consists of 100 control steps, i.e., $T=100$ in Eq~\eqref{energy_cost}.  

2) The 3D system is defined as $ \dot{x} = y + 0.5z^2, \dot{y} = z  ,\dot{z} = u$,
where system state $s=(x(t), y(t), z(t))^{'}$, $X=X_{0}=[-0.5, 0.5]^{3}$, $u(t) \in U = [-10, 10]$, and $T=100$.  A sampling period $\tau = 0.05$ is used to discretize the ordinary differential equations (ODEs) into a discrete system.

3) The cartpole system is described as

\noindent\begin{minipage}{.5\linewidth}
\begin{displaymath}
  \begin{cases}
s_1(t+1) = s_1(t)+\tau s_2(t)\\
s_2(t+1) = s_2(t) + \tau s_{acc} \\
s_3(t+1) = s_3(t) + \tau s_4(t) \\ 
s_4(t+1) = s_4(t) + \tau \theta_{acc}
\end{cases}
\end{displaymath}
\end{minipage}%
\begin{minipage}{0.5\linewidth}
\begin{displaymath}
  \begin{cases}
\psi = \FS{u+m_p l s_{4}^2\sin{s_3}}{m_t}\\
\theta_{acc} = \FS{(g\sin{s_3}-\cos{s_3}\psi)m_t}{l(1.333 - m_{p}(\cos{s_3})^{2})} \\
s_{acc} = \FS{\psi - m_p l  \cos{s_3} \theta_{acc}}{m_t}
\end{cases}
\end{displaymath}
\end{minipage}
with $m_{c}=1, m_{p}=0.1, m_{t} = 1.1, g=9.8, l=1$, $\tau = 0.02$, $T=200$ and $s=(s_1, s_2, s_3, s_4)^{'}$. $X = \{ s | \ s_1 \in [-2.4, 2.4], s_3 \in  [-0.209, 0.209]\}$ and $X_{0} = [-0.2, 0.2]^{4}$ ($X_0 \subset X$ for further reachability analysis).

In our testing for each example, we randomly sample 500 initial system states from $X_{0}$, and compare the results from our \toolname \ framework and other baselines.  The comparison on control robustness and energy efficiency is based on simulations within a Python environment that we developed. The further analysis on verifiability, with safety consideration, is done via formal analysis as outlined in Section~\ref{verification}. The training and testing, including the recording of verification time, are conducted on a server with 4-core 3.60GHz Intel Core and NVIDIA GTX TITAN. 


\begin{table}
\centering
  \begin{tabular}{cccccccl}
    \hline
    Oscillator & $\kappa_1$ & $\kappa_2$ & $A_S$\cite{wang2020energy} & $A_{W}$ & $\kappa_{D}$ & $ \kappa^{*}$ \\
    \hline
    $S_r$ (\%) & 85  & 79.4  & 88.4  & 98 & 98.4 & \textbf{98.8}\\
    $e$  &  94.1 & 97.9 & 94.2 & 96.3 & 94.6 & \textbf{86.2} \\
    $L$ & 35.4 & 15.1  & -  & - & 20.5 & \textbf{7.6} \\
   \hline
    3D system & & & & & & \\
   \hline
     $S_r$ (\%) & 91 & 88.6  & 96.8  & 98.2  & 97.6  & \textbf{99} \\
    $e$ & 16.6 & 16.6  &  13.5  & 12.7 & 12.3 & \textbf{11.8}\\
    $L$ & 251 & \textbf{0.72}  & -  & - & 12.1  & 7.1 \\
    \hline
    Cartpole & & & & & & \\
    \hline
    $S_r$ (\%) & 81.6 & 84 & 90.4  & \textbf{99}  & \textbf{99}  & 98.6\\
    $e$ &  106.1 & 74.7  & 84.8  &  28.8 & 29 & \textbf{27.7} \\
    $L$  &  359.7  & 303.9   & -  & - & 126.1  & \textbf{72.5} \\
    \hline
\end{tabular}
\caption{\MakeTextUppercase{C}omparison of \toolname \ with baselines. $S_a$ is the safe control rate without attacks or measurement noises to the system state yet, $e$ is the control energy consumption, and $L$ is the \MakeTextUppercase{L}ipschitz constant. \MakeTextUppercase{T}he baselines include $\kappa_1 \ only$, $\kappa_2 \ only$, switching adaptation method $A_S$, intermediate mixed controller $A_{W}$ after adaptive mixing in \toolname \ (no distillation), and direct distillation result $\kappa_D$ from $A_W$ (no consideration of robustness). $\kappa_2$ in the 3\MakeTextUppercase{D} system is a polynomial controller~\cite{sassi2017linear} and has a very small $L$. \MakeTextUppercase{T}he \MakeTextUppercase{L}ipschitz constant for $A_S$ and $A_W$ cannot be measured and thus are denoted as '-'. \MakeTextUppercase{W}e can see the significant improvement from our approach.}
\label{results_table}
\end{table}

\smallskip
\noindent
\textbf{Effectiveness of our approach over baselines:} We compare the following methods to demonstrate the effectiveness of our approach: 1) using a single control expert, e.g., \textit{$\kappa_{1}$ only} or \textit{$\kappa_{2}$ only}; 2) a state-of-the-art switching adaptation control method from~\cite{wang2020energy}, denoted as \textit{$A_S$}; 3) the intermediate mixed controlled design (i.e., before distillation) in \toolname, denoted as \textit{$A_W$}; 4) the direct distillation result from $A_W$ without any adversarial training and regulation, denoted as $\kappa_{D}$; and 5) the robust distillation result from $A_W$, which is what our \toolname \ eventually produces, denoted as $\kappa^{*}$. 

The comparison results are shown in Table~\ref{results_table}. We can see that compared with $\kappa_1$, $\kappa_2$ and $A_S$ (single expert or switching adaptation method), $\kappa^{*}$ obtained from our \toolname \ framework provides significant improvement on the safe control rate (without attacks or measurement noises to the system state yet) and control energy efficiency. 
Compared with the intermediate mixed controller design \textit{$A_W$} and the direct distillation result $\kappa_{D}$, $\kappa^{*}$ is easier to verify with the smaller Lipschitz constant (more about this later; note that the mixed controller design cannot be verified with current tools and does not have associated Lipshitz constant). Our approach also has smaller control energy consumption than \textit{$A_W$} and $\kappa_{D}$.


\begin{table}
\centering
  \begin{tabular}{|c|cc|cc|l}
    \hline
    & \multicolumn{2}{|c|}{Under adversarial attacks}  & \multicolumn{2}{|c|}{With measurement noises} \\
    \hline
    Oscillator &  $\kappa_{D}$ & $\kappa^{*}$ & $\kappa_{D}$ & $\kappa^{*}$  \\
    \hline
    $S_r$ (\%) & 95.2  & \textbf{98.8} & 98.4 & \textbf{98.8}   \\
    $e$ & 837.3 & \textbf{132.1} & 383.8  & \textbf{98.9}  \\ 
    \hline
    3D system &  $\kappa_{D}$ & $\kappa^{*}$ & $\kappa_{D}$ & $\kappa^{*}$  \\
    \hline
    $S_r$ (\%) & 91.6  & \textbf{98.2} & 96 & \textbf{98.8}  \\
    $e$ & 149.2 & \textbf{25.7} & 61.3 & \textbf{15.5} \\
    \hline
    Cartpole &  $\kappa_{D}$ & $\kappa^{*}$ & $\kappa_{D}$ & $\kappa^{*}$ \\
    \hline
    $S_r$ (\%) & 92.2  & \textbf{96} & 96.4  &  \textbf{98.4}  \\
    $e$ & 30.6  & \textbf{29.1} & 31.1  & \textbf{28.1} \\
    \hline
\end{tabular}
\caption{\MakeTextUppercase{C}omparison of $\kappa^{*}$ and $\kappa_D$ under optimized adversarial attacks and measurement noises to the system state. $\kappa^{*}$(\toolname) shows stronger robustness, indicating the efficacy  of our robust distillation design. \MakeTextUppercase{N}ote that while not shown in the table, $A_W$ performs slightly worse than $\kappa^{*}$ in energy efficiency, and other baselines perform much worse in both robustness and energy efficiency.}
\vspace{-12pt}
\label{robust_table}
\end{table}



\begin{figure*}
    \centering
    \includegraphics[width=0.85\linewidth]{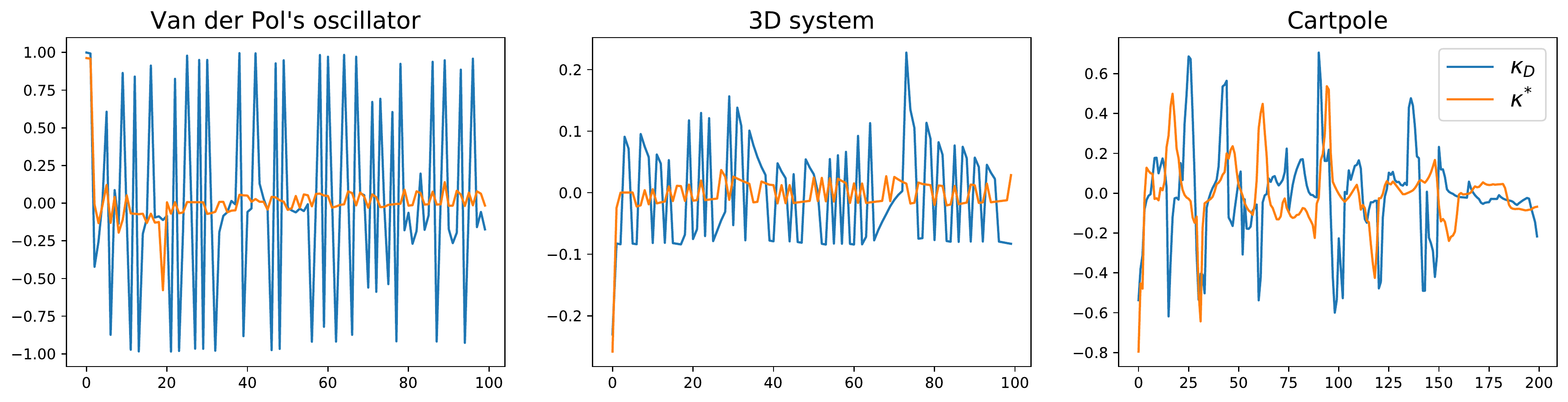}
    \caption{The normalized control input signal $u(t)$ when the system encounters adversarial attacks. Compared with $\kappa_{D}$, $\kappa^{*}$ obtained from \toolname \ is more robust to these attacks and consumes much less energy, indicating the effectiveness of our robust distillation design. Note that the performance difference between $\kappa^{*}$ and $\kappa_{D}$ in cartpole is less significant than the others because cartpole is an unstable system. }
    \label{fig_attack_u}
    \vspace{-18pt}
\end{figure*}

\begin{figure}
    \centering
    \includegraphics[width=0.85\linewidth]{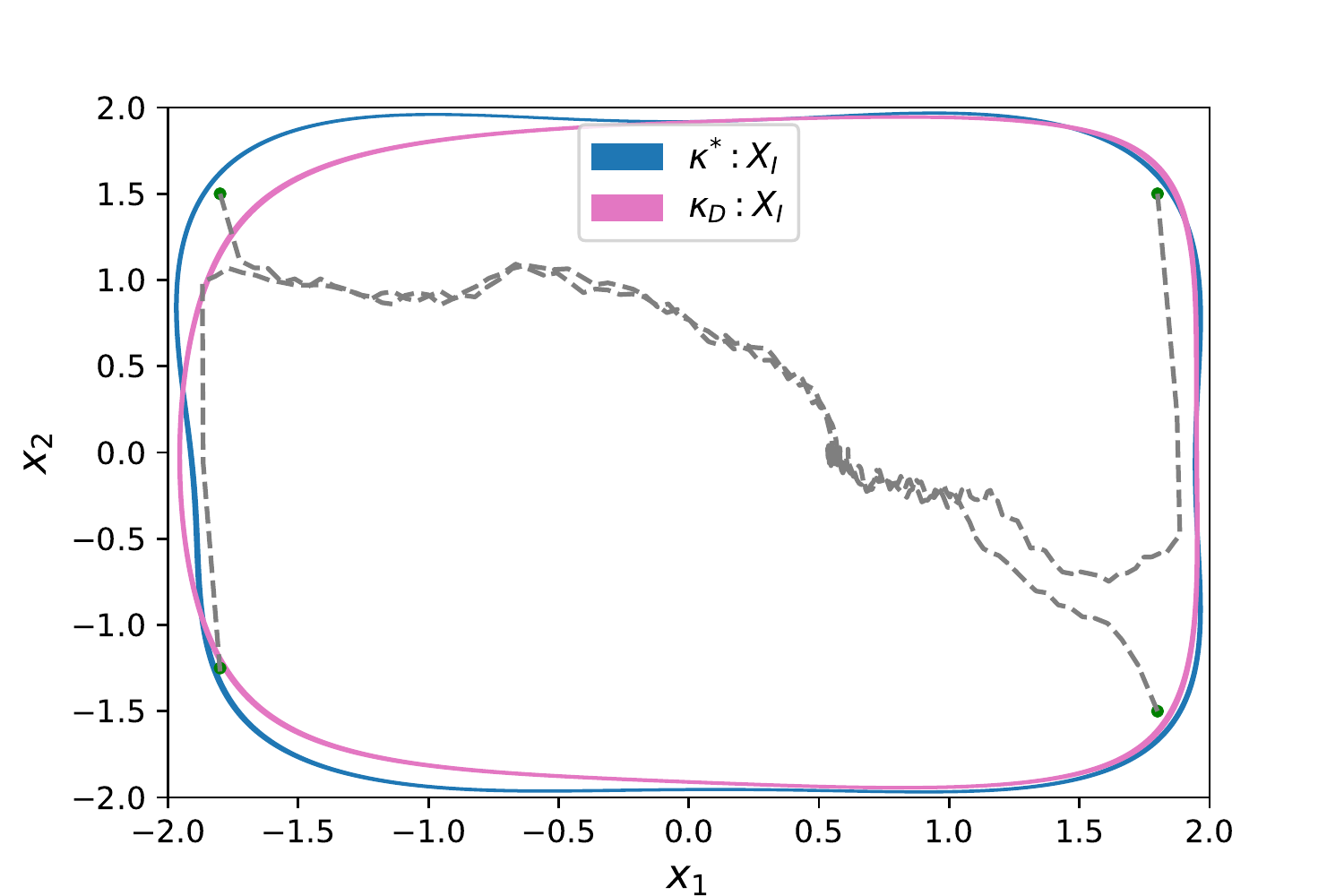}
    \caption{Invariant set $X_{I}$ of the oscillator system for $\kappa^{*}$ and for $\kappa_{D}$. 
    Although with higher order Bernstein polynomials, the $X_{I}$ for $\kappa_D$ is more conservative than the one for $\kappa^{*}$ due to its slightly larger approximation error bound $\epsilon$. $X_{I}$ for $\kappa^{*}$ is computed using the approach from~\cite{xue2018robust} (as stated in Section~\ref{verification}) in about 32 minutes, while needs around 11 hours for $\kappa_D$, showing the effectiveness of our approach in reducing verification time (improving verifiability). We also conducted 1500 simulations for different initial states within $X_{I}$ for $\kappa^{*}$, and as expected, all the trajectories including the 4 dashed lines shown in the figure are indeed safe (the green dots are initial states; all 4 trajectories eventually are stable at around $(0.5, 0)$). 
    }
    \label{os_inner}
    \vspace{-12pt}
\end{figure}

\begin{figure}
    \centering
    \includegraphics[width=0.85\linewidth]{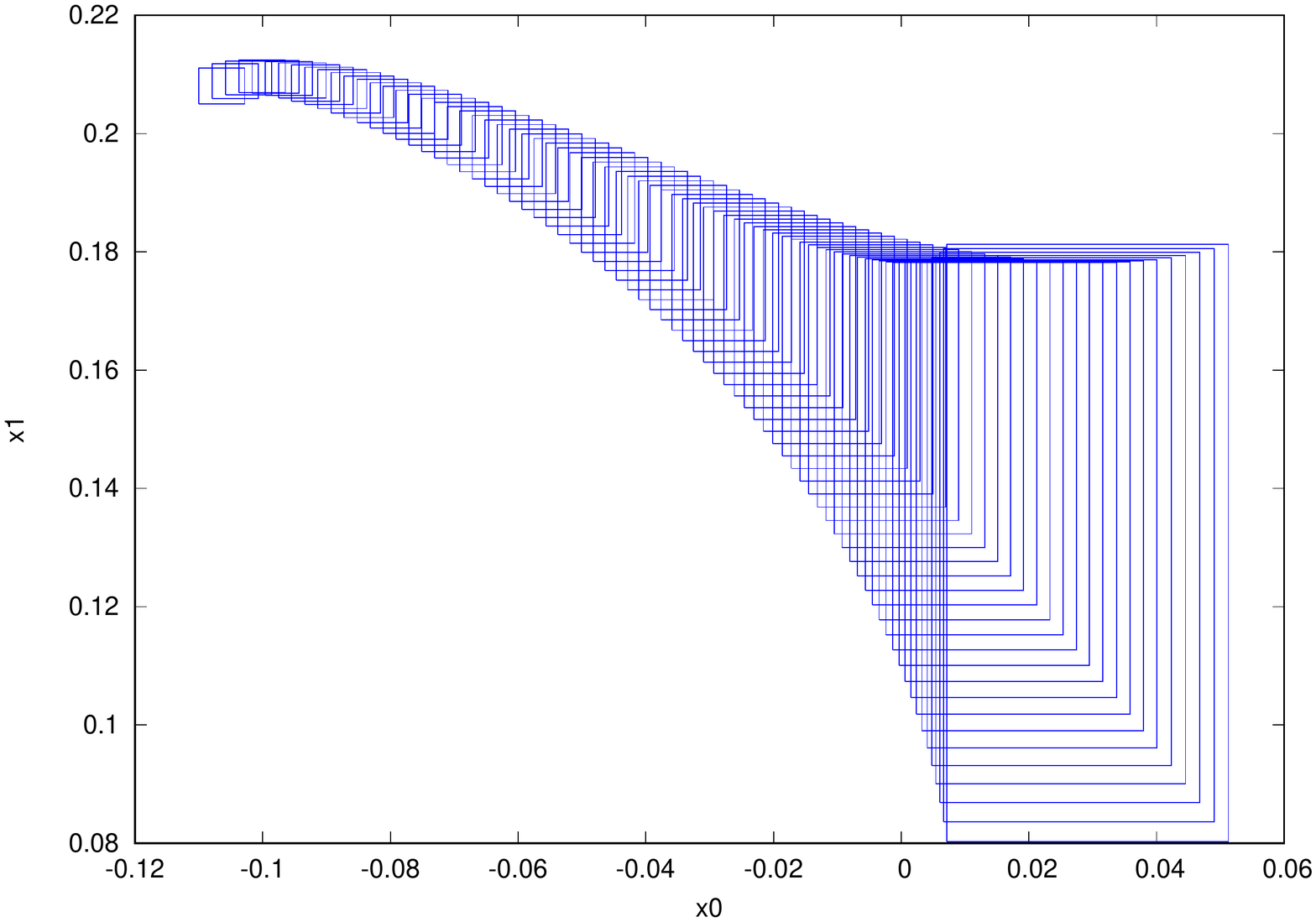}
    \caption{Reachable set in the 3D system within the first 15 control steps from the initial state set on the upper left corner, as $s=(x, y, z)\in [-0.11, -0.105]\times[0.205, 0.21]\times[0.1, 0.11]$. Note that only $(x, y)$ is plotted. We show this system because the difference between its Lipschitz constants of $\kappa_{D}$ and $\kappa^{*}$ is the smallest in all three examples. Nevertheless, $\kappa_{D}$ cannot be verified because of a memory segmentation fault after 12 reachable set computation, caused by its large Lipschitz constant. In contrast for $\kappa^{*}$, the verification can be successfully completed within minutes (the reachable set does not go out of $X$ and the system is verified to be $Safe$).}
    \label{reachable_set}
\end{figure}

\smallskip
\noindent
\textbf{Further analysis on robustness and verifiability:} 
We then further tested the effectiveness of our approach in improving control robustness and system verifiability, considering the cases where the system encounters adversarial attacks or measurement noises to the system state. 
Specifically, the measurement noise is a random variable sampled from an uniform distribution and added to the system state $s(t)$ at every step. The adversarial attack is generated by FGSM with a bound that is the same or larger than the one assumed in our robust distillation. In the experiments, the noises and the attacks are between $10\%-15\%$ of the system state value bound. 
Table~\ref{robust_table} shows the result comparison between our approach (generating $\kappa^{*}$) and direct distillation (generating $\kappa_D$).   
We can see that our approach benefits from the probabilistic adversarial training and robust distillation design, producing results that are more robust with respect to the adversarial attacks and measurement noises, as well as have smaller energy consumption.  
The control signal (and its energy consumption) from our results is also more stable under attacks, which is visualized in Fig.~\ref{fig_attack_u}. Furthermore, we conducted formal analysis of the system properties (i.e., computing invariant set and conducting reachability analysis for safety verification) for the oscillator and the 3D system, respectively, as shown in Figs.~\ref{os_inner} and~\ref{reachable_set}. The results demonstrate the effectiveness of our \toolname \ in reducing verification time. 

\section{Conclusion}\label{conclusion}
In this paper, we propose a novel framework \toolname \ to automatically learn an improved neural network controller from multiple control experts for LE-CPSs. Our approach first learns a system-level adaptive mixing strategy with optimal weights dynamically assigned to the experts using reinforcement learning, and then synthesize a single student neural network controller with robust distillation. Experiments demonstrate that our approach can significantly improve system control robustness, control energy efficiency, and verifiability.


\bibliographystyle{IEEEtran}
\bibliography{IEEEabrv,main}

\end{document}